\documentclass{moriond}

\usepackage{amsmath}
\usepackage{amssymb}
\usepackage{xcolor}

\def\be{\begin{equation}}
\def\ee{\end{equation}}
\def\bea{\begin{eqnarray}}
\def\eea{\end{eqnarray}}

\newcommand{\Photo}{}

\begin{document}
\begin{flushright}
\begin{tabular}{l}
{\tt \footnotesize IPPP/17/54}\\
\end{tabular}
\end{flushright}

\vspace*{4cm}
\title{SEARCHING FOR HEAVY STERILE NEUTRINOS IN KAON DECAYS}

\author{ C. WEILAND }

\address{Institute for Particle Physics Phenomenology, Department of Physics,\\ 
Durham University, South Road, Durham DH1 3LE, United Kingdom}

\maketitle\abstracts{We present here a study of the impact of a heavy neutrino (or heavy neutral lepton) on leptonic and semileptonic kaon decays.
We used a simplified model consisting of 3 light neutrinos responsible for neutrino oscillations and a heavy sterile neutrino. We found that it can lead
to large deviations from the Standard Model predictions for leptonic decays, in conflict with experimental measurements of $K_{e2}$ and the lepton universality test $R_K$. This allows to derive
new constraints on the leptonic mixing for heavy sterile neutrinos. No tension was found when considering the semileptonic decays. Finally, we point out the potential of the 
decay $K_L \rightarrow \nu \nu$ as a clear signature of physics beyond the Standard Model.}

\section{Introduction}

The observation of neutrino oscillations provides a clear evidence of the existence of physics beyond the Standard Model (SM).
Among the many extensions of the SM that were introduced to explain this phenomenon and generate neutrino masses and mixing,
one of the simplest is the addition of sterile neutrinos. Being fermionic gauge singlets, they can have a Majorana mass term whose value is not necessarily related to the typical
mass scales of the SM like the QCD scale or the electroweak scale. Depending on their mass, sterile neutrinos can have very different yet observable effects. 
For example, eV-scale sterile neutrinos could 
solve neutrino oscillation
anomalies~\cite{Mueller:2011nm,Huber:2011wv,Mention:2011rk,Aguilar-Arevalo:2013pmq,Giunti:2010zu}, while keV-scale sterile neutrinos
can be dark matter candidates~\cite{Asaka:2005an,Abada:2014zra}. Above $10^9$ GeV, they could explain the observed baryonic asymmetry of the Universe through high-scale
leptogenesis~\cite{Fukugita:1986hr,Davidson:2008bu}. Sterile
neutrinos in the range MeV--GeV have been introduced in minimal models like the $\nu$MSM~\cite{Asaka:2005an,Asaka:2005pn} and lead to visible
effects in meson decays~\cite{Shrock:1980vy,Shrock:1980ct,Abada:2012mc,Abada:2013aba}.

We present here results of a study~\cite{Abada:2016plb} where we focused in particular on the impact of heavy sterile neutrinos on rare kaon leptonic and semileptonic decays
with final state neutrinos. These charged kaon decays are searched for at NA62~\cite{Hahn:1404985,Volpe:2017lyp} while neutral kaon decays are studied by KOTO~\cite{Ahn:2016kja}
and NA64~\cite{Gninenko:2013rka,Andreas:2013lya,Gninenko:2016rjm}. Our study can be useful as well for the TREK/E36 
experiment at J-PARC, where the data analysis is currently under way~\cite{Bianchin:2016vds}. It will further test the lepton universality 
in kaon two-body decays ($K_{\ell2}$) and search for a heavy neutrino~\cite{Kohl:2013rma}.

\section{A simplified model}

In order to capture the generic effects due to the presence of heavy neutrinos, we use a simplified $3+1$ model with Majorana neutrinos. Three of them are light and 
responsible for neutrino oscillations while the fourth is heavier, its mass being a free parameter in our study. An immediate consequence is the 
modification of the charged and neutral currents as
\begin{align}
\label{mod.current} \mathcal{L}_{W^\pm} &\supset
-\frac{g_2}{\sqrt{2}} \, W^-_\mu \, \, \bar
\ell_\alpha\,U_{\alpha i} \gamma^\mu\, P_L\, \nu_i\ , \\\nonumber
\mathcal{L}_{Z}&\supset -\frac{g_2}{4 \cos \theta_W} \, Z_\mu \,
 \bar \nu_i \,\gamma ^\mu \left[ P_L\,( U^\dagger
  U)_{ij} - P_R \, ( U^\dagger U)_{ij}^* \right] \nu_j\ , 
\end{align}
where $g_2$ is the weak coupling constant, $\theta_W$ the weak mixing angle and $U$ is a $3\times4$ mixing matrix. Being rectangular, $U$ is obviously non-unitary but it nonetheless verifies 
$\sum_{i=1}^4 |U_{\alpha i}|^2=1$. It arises from the diagonalisation of the neutrino and charged lepton mass matrices and is defined as
\begin{equation}
\label{Udef}
U_{\alpha i}\, =\,\sum_{k=1}^3 V^*_{k\alpha}\, U_{\nu_{ki}}\, ,
\end{equation}
with
\begin{equation}
\label{RotDiag}
\ell'_L =\, V\, \ell_L \,,\qquad  \nu'_L \,=\, U_\nu\, \nu_L\, .
\end{equation}
where $\ell'$ and $\nu'$ are the weak gauge eigenstates while $\ell$ and $\nu$ are the mass eigenstates, $V$ and $U_\nu$ being unitary matrices. Thus, this
modified lepton mixing matrix depends on the active-sterile mixing angles which are free parameters of our study as well.

\section{Scan of the parameter space}
\label{sec:scan}

Our study has 8 free parameters: the mass of the lightest neutrino $m_1$, the mass of the heavy neutrino $m_4$, the new active-sterile mixing angles
$\theta_{14},\, \theta_{24},\, \theta_{34}$ used to build $U$ according to Eq.~4 of our study~\cite{Abada:2016plb} as well as the Dirac \textit{CP}-violating phases
$\delta_{13},\, \delta_{41},\, \delta_{43}$. We have explicitly checked that the three Majorana
phases which are present as well do not significantly affect our result, either cancelling or giving a contribution suppressed by the light neutrino masses. The light neutrino mass differences
and mixing are chosen according to the best fit point of a recent global fit to neutrino oscillation data~\cite{Esteban:2016qun}. We performed three random scans of the parameter space,
using flat priors on the Dirac \textit{CP} phases and logarithmic priors on all other scan parameters, with the size of our samples and the ranges considered given in Table~\ref{tab:scan}.
\begin{table}[t]
  \begin{center}
    \begin{tabular}{|l|c|c|}
     \hline
	 & Sample 1 & Sample 2 \\
     \hline
	Number of points 				& 200000 		& 40000 \\
	$m_1\, (\mathrm{eV})$  				& $[10^{-21};\,1]$ 	& $[10^{-21};\,1]$ \\
	$m_4\, (\mathrm{GeV})$  			& $[0.1;\, 1]$		& $[0.27;\,0.35]$ \\
	$\theta_{14},\, \theta_{24}$ 			& $[10^{-6};\,2 \pi]$	& $[10^{-6};\,2 \pi]$ \\
	$\theta_{34}$					& $[10^{-6};\,2 \pi]$	& $[0.1;\,2 \pi]$ \\
	$\delta_{13},\delta_{41},\, \delta_{43}$	& $[0;\, 2 \pi]$	& $[0;\, 2 \pi]$ \\
     \hline
    \end{tabular}
    \caption{\label{tab:scan} Input parameters in our random scan. Both samples were combined for this study.}
 \end{center}
\end{table}

\begin{figure}[ht!]
\centering
\includegraphics[width=0.33\linewidth]{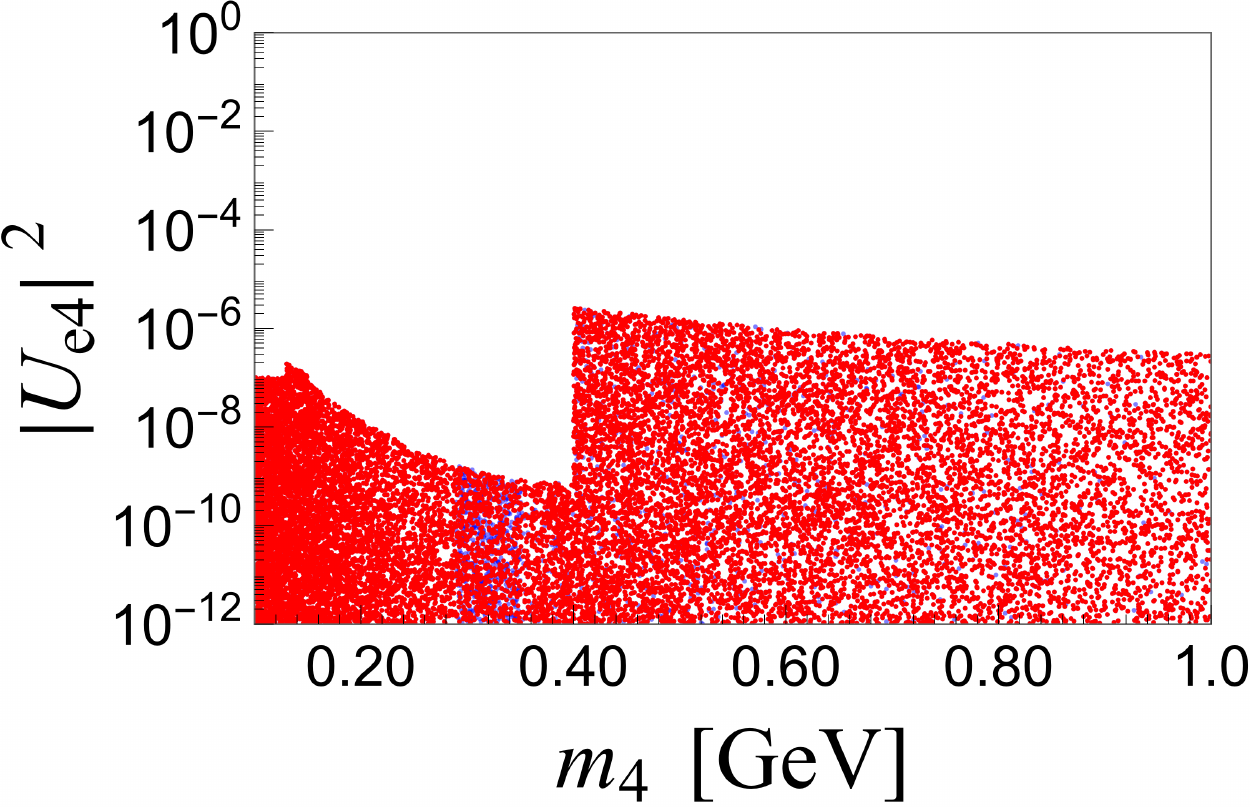}~\includegraphics[width=0.33\linewidth]{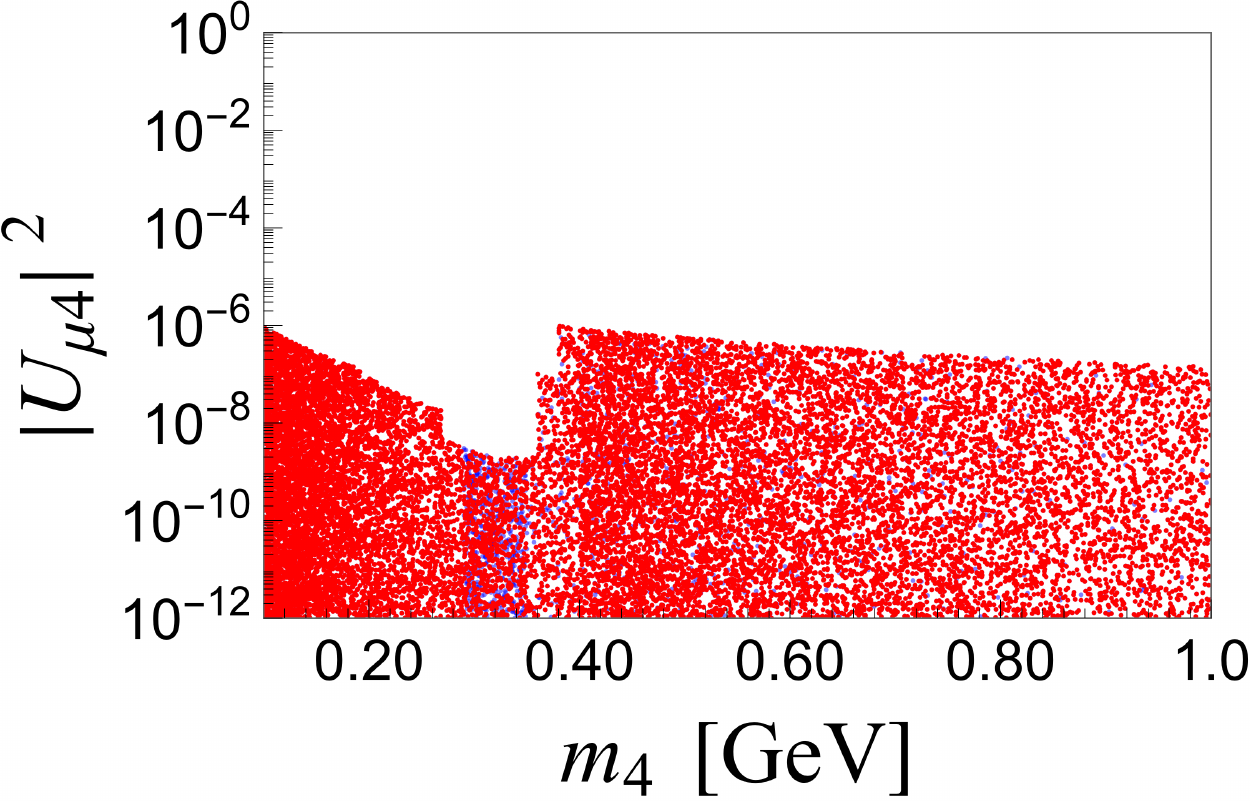}~\includegraphics[width=0.33\linewidth]{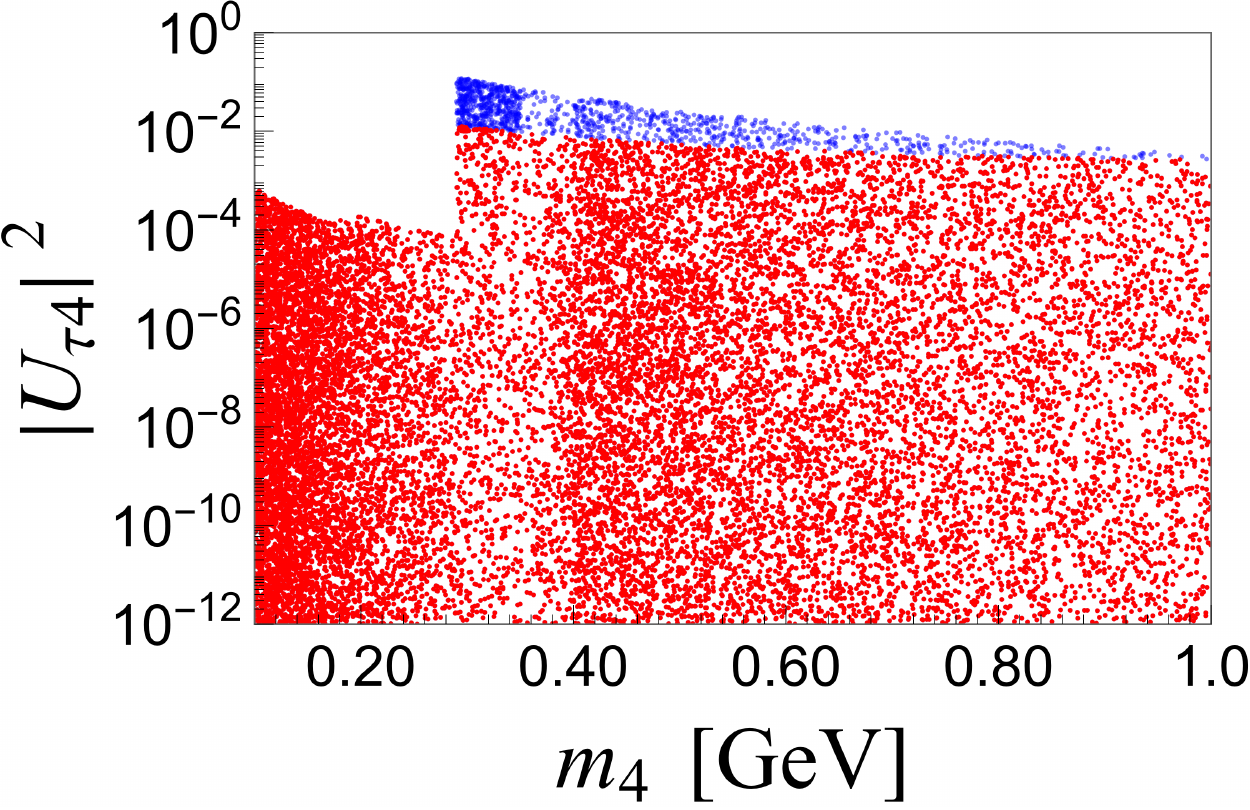}
\caption{\small \sl Points of our random scan remaining after applying the experimental constraints. The red points agree with all constraints while the blue ones are
  excluded by the leptonic $\tau$ decays.}
\label{fig:1}
\end{figure}

Focusing on the region where the heavy neutrino mass is between $0.1$ and 1~GeV, we need to include
experimental constraints on the active-sterile mixing. First, we consider the limits coming from the direct searches~\cite{Atre:2009rg} which give the strongest constraints 
for most of the parameter space. Then, we include constraints from both radiative lepton flavour violating decays and 3-body decays, comparing our predictions
based on the formulas derived by Ilakovac \textit{et al.}~\cite{Ilakovac:1994kj} with the MEG~\cite{TheMEG:2016wtm}, SINDRUM~\cite{Bellgardt:1987du}, Belle~\cite{Hayasaka:2010np}
and BaBar~\cite{Aubert:2009ag} results. We also compare our prediction of $W \rightarrow \ell \nu$, $Z \to \nu \nu$ and $\tau \to \ell \nu\nu$ ($\ell = e,\mu$), which agree with a previous
independent calculation~\cite{Abada:2013aba}, with LHCb~\cite{Aaij:2016qqz} and LEP measurements~\cite{Olive:2016xmw}. We include as well lepton universality tests in
pion decays~\cite{Abada:2013aba}. Finally, we check that for the model used in this study and after applying all other constraints the Fermi constant $G_F$ is to an excellent approximation
equal to $G_\mu$, its value extracted from muon decays. Points surviving constraints are presented in Fig.~\ref{fig:1}. 

\section{Numerical results}

In this work, we focused on processes for which hadronic theoretical uncertainties are under control and which contain final state on-shell neutrinos, namely $K_{\ell 2}$, $K_{\ell 3}$,
$K\to \pi \nu \nu$ and $K_L\to \nu\nu$. Detailed formulas including massive neutrinos can be found in our main article~\cite{Abada:2016plb}.

\subsection{Leptonic decays, $K_{\ell 2}$}
To better understand the impact of a heavy neutrino on these decays, we will first present analytical formulas for $\mathrm{Br} (K\to \ell \nu)$ and highlight the difference with the SM.
Shrock~\cite{Shrock:1980vy,Shrock:1980ct} first pointed out the impact of sterile neutrinos on kaon and pion decays and their use to put bounds on neutrino masses and lepton mixing matrix elements.
Since the heavy neutrino escapes the detector unobserved, one has to sum over all kinematically accessible neutrinos, whose number is denoted here by N, giving the expression
\begin{align}\label{eq:BrKl2}
\mathrm{Br} (K\rightarrow \ell \nu )=\frac{G_F^2\tau_K}{8\pi m_K^3}  |V_{us}|^2 f_K^2  
\sum_{i=1}^N |U_{\ell i}|^2 \lambda^{1/2}(m_K^2,m_\ell^2,m_{\nu_i}^2)   \left[ m_K^2 (m_\ell^2 +m_{\nu_i}^2) - (m_\ell^2 -m_{\nu_i}^2)^2\right] \,,
\end{align}
where $\lambda$ is the K\"all\'en function, $\tau_K$ is the kaon lifetime, $f_K$ is the kaon decay constant and $V$ is the $CKM$ matrix. First, the presence of a heavy neutrino modifies $U$ such that
$\sum_{i=1}^3 |U_{\ell i}|^2 < 1$ which leads to non-unitarity effects for the light neutrino contributions, even when the heavy neutrino is not kinematically accessible. Second, the decay 
$K_{\ell 2}$ is helicity suppressed and the presence of a heavy neutrino in the final state would lift this helicity suppression, increasing the corresponding partial width. The above formula can be extended to other leptonic decays of
pseudoscalar mesons by substituting $\tau_K$,  $f_K$ and $V_{us}$ with the corresponding parameters. 

\begin{figure}[ht]
\centering
\hspace*{-6mm}\includegraphics[width=0.5\linewidth]{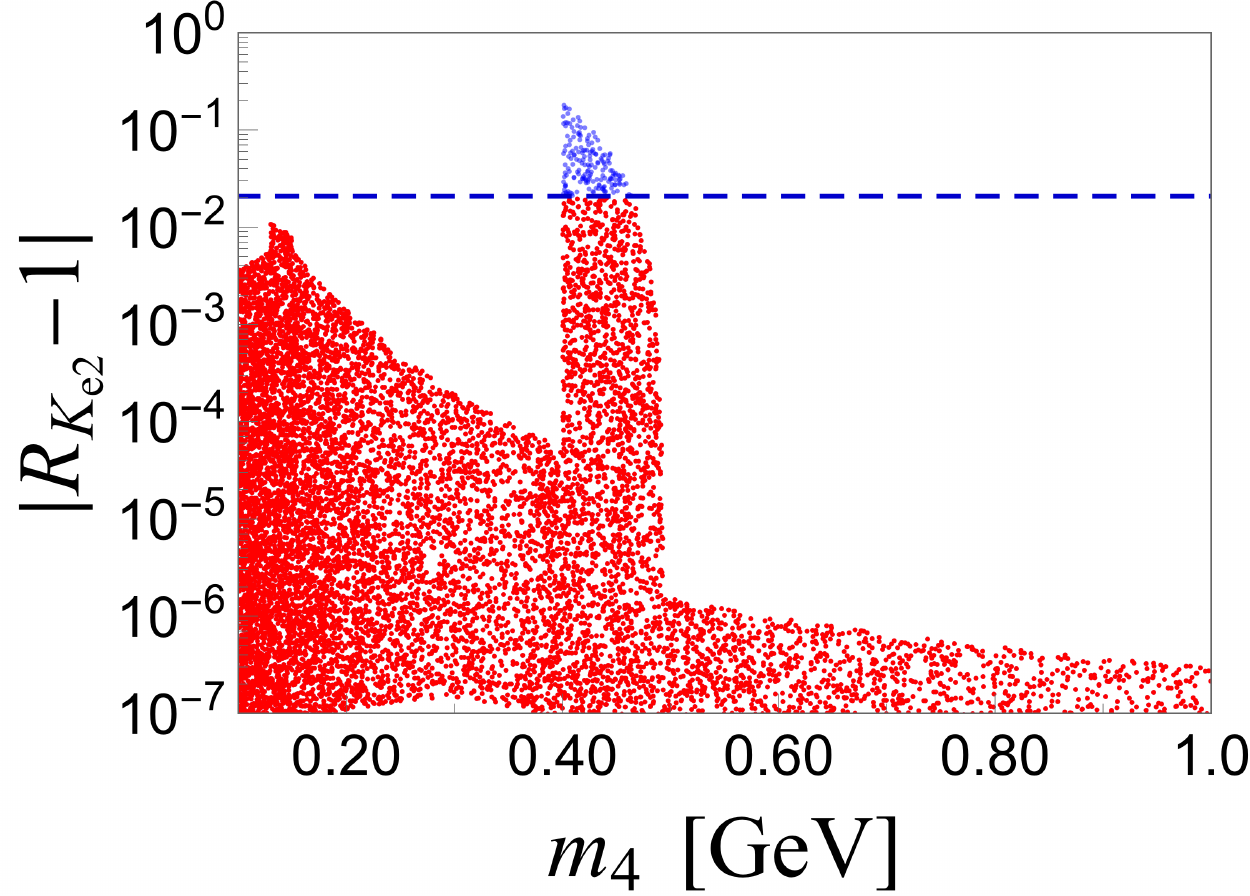}~\includegraphics[width=0.5\linewidth]{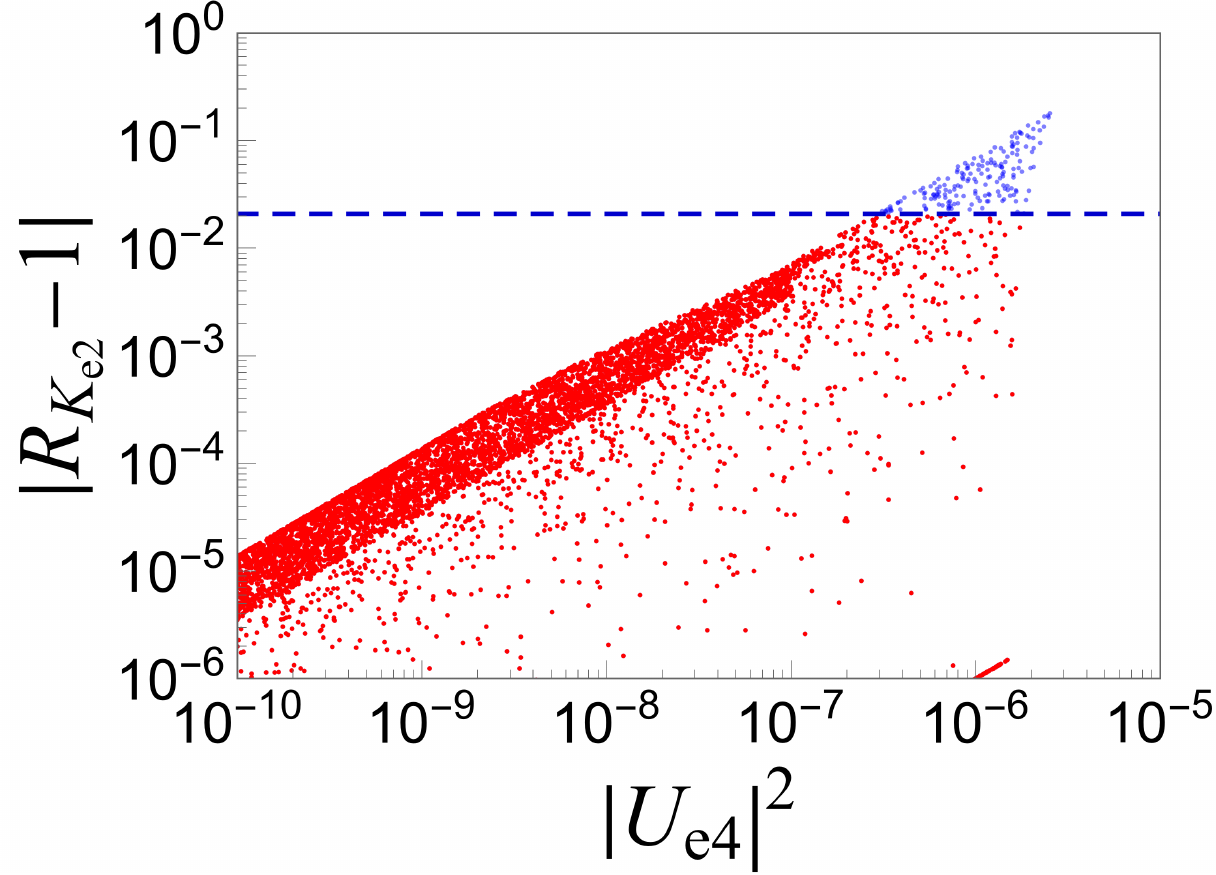}
\caption{\small \sl   $|R_{Ke2}-1|$ as a function of the sterile neutrino mass $m_4$ (left) and of the leptonic mixing $U_{e4}$ (right). The red points agree with all constraints
	while the blue ones are in conflict with $R_{Ke2}^\mathrm{exp}$ shown by the dashed line.}
\label{fig:2}
\end{figure}

\begin{figure}[ht]
\centering
\hspace*{-6mm}\includegraphics[width=0.65\linewidth]{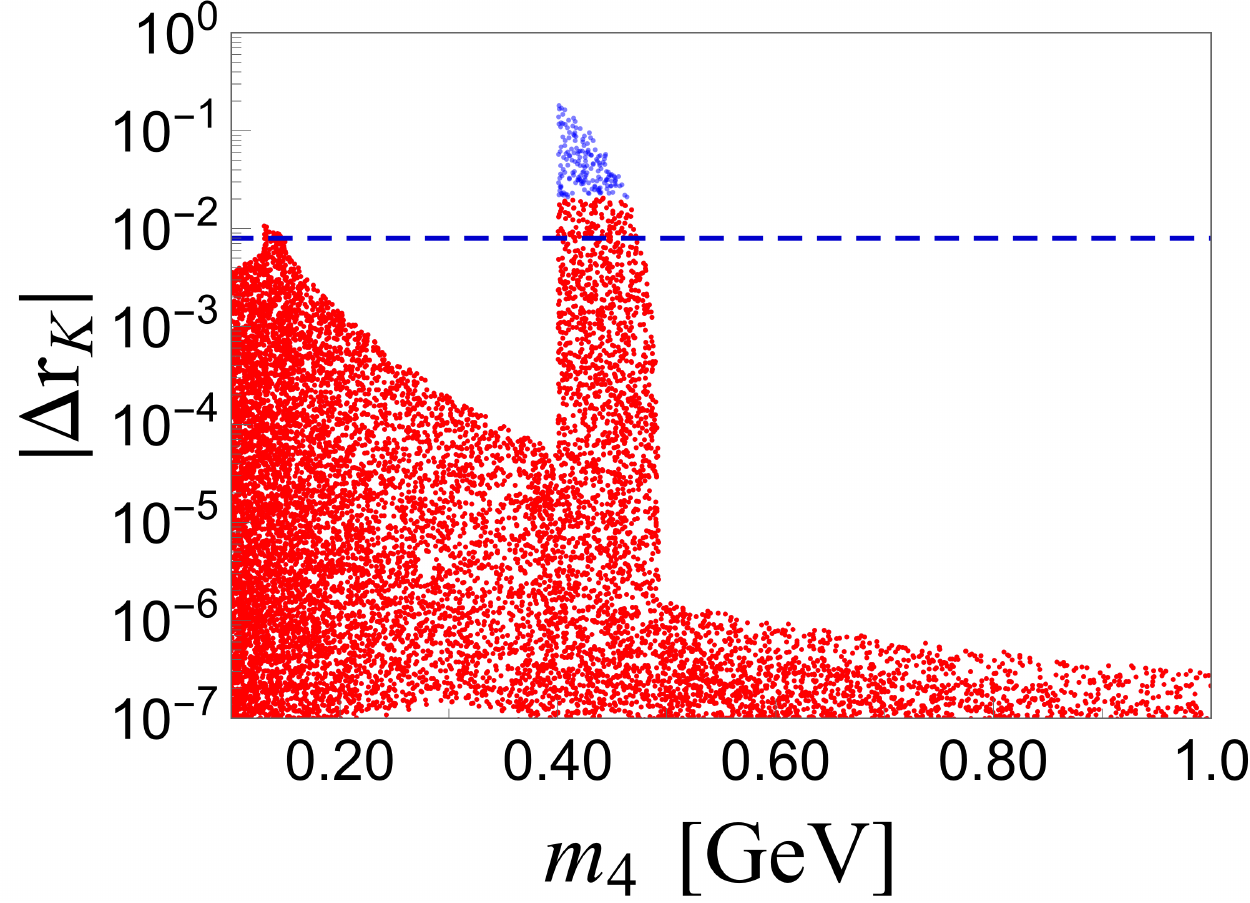}
\caption{\small \sl $\Delta r_K$ as a function of the sterile neutrino mass $m_4$. The red points agree with all constraints
	while the blue ones are in conflict with $R_{Ke2}^\mathrm{exp}$.  The dashed blue line corresponds to the maximally allowed deviation from the experimental 
measurement.}
\label{fig:2bis}
\end{figure}

Nowadays, lattice computations of $f_K$, and especially of $f_K/f_\pi$, have reached a
sub-percent accuracy~\cite{Aoki:2016frl}, allowing to distinguish new physics effects at the percent level or smaller from the SM prediction. Our predictions are presented in Fig.~\ref{fig:2} 
for the decay $K\rightarrow e \nu$, where we used the ratio
\begin{equation}
R_\mathcal{O}= \frac{\mathcal{O}}{\mathcal{O}^\mathrm{SM}}\,,
\end{equation}
of an observable $\mathcal{O}$ to its SM value. We can see that a heavy neutrino with a mass between $400~\mathrm{MeV}$ and $m_K$ can induce deviations at the percent-level while being agreement 
with all other experimental constraints. This demonstrates the potential of $K_{e 2}$ to provide additional constraints due to the partial lifting of the helicity suppression coming from
an extra sterile neutrino. This was confirmed by investigating the $K_{\mu 2}$ decay, where the helicity suppression is weaker and the maximal deviation consequently smaller. However, the ratio 
$\Delta r_K=R_{Ke2}/ R_{K\mu2}-1$ allows to derive even stronger bounds as can be seen in Fig.~\ref{fig:2bis}
since it has smaller theoretical and experimental uncertainties. This will be even 
more relevant in the future because of the reduction of the experimental uncertainty by a factor of $\sim2$ expected by the TREK experiment~\cite{Kohl:2013rma,Bianchin:2016vds}.

\subsection{Semileptonic decays, $K_{\ell 3}$}

\begin{figure}[t!]
\centering
\hspace*{-6mm}\includegraphics[width=0.5\linewidth]{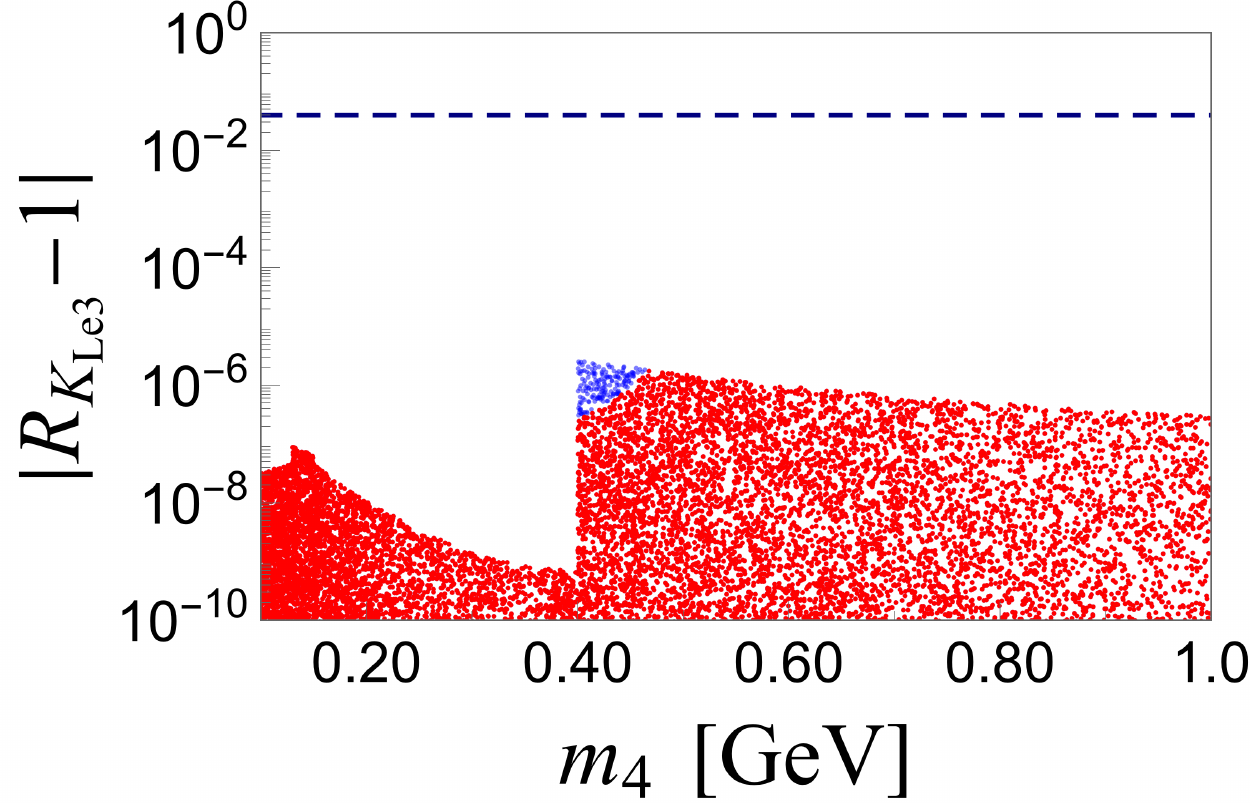}~\includegraphics[width=0.5\linewidth]{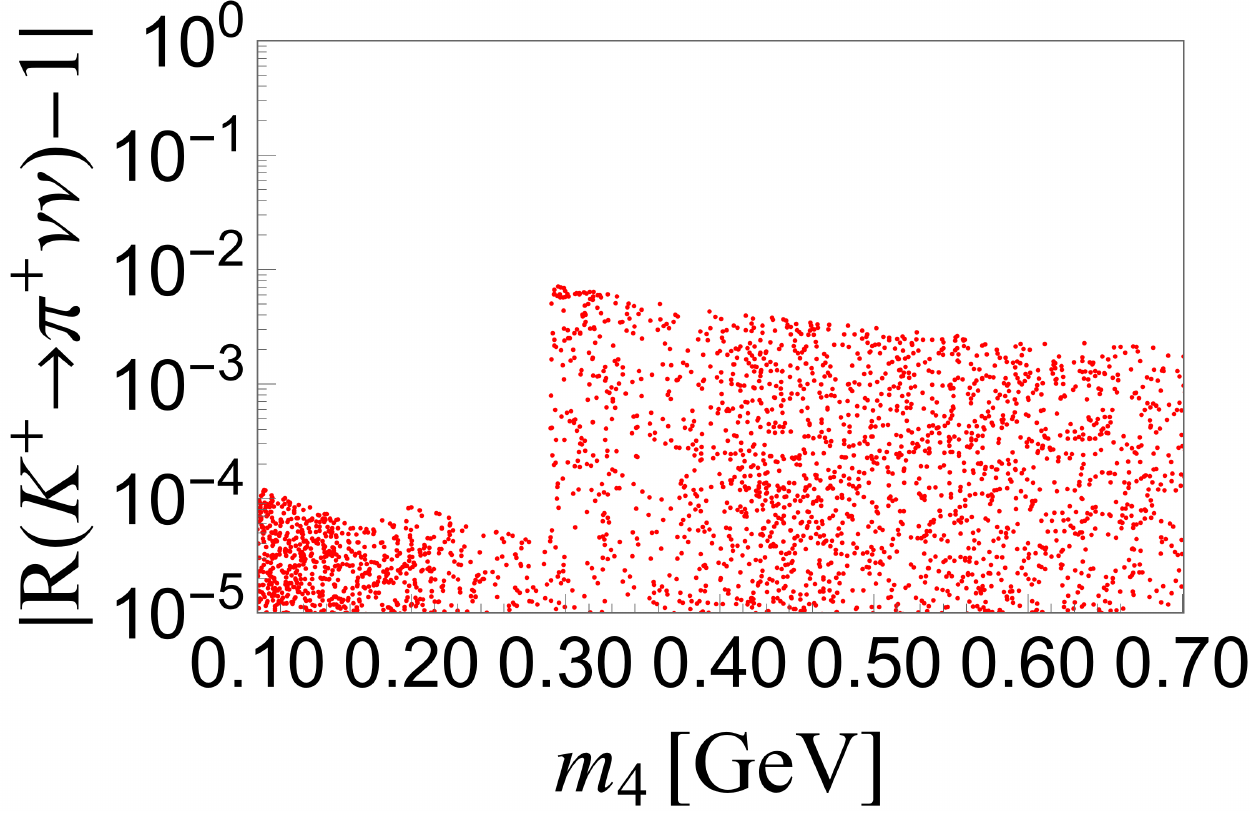}
\caption{\small \sl  Predictions for $|R_{KLe3}-1|$ and $|R(K^+\rightarrow\pi^+\nu\nu)-1|$ as functions of the mass of the sterile neutrino $m_4$. 
The red points agree with all constraints while the blue ones are in conflict with $R_{Ke 2}^\mathrm{exp}$. The dashed blue line corresponds to the maximally allowed deviation from the experimental 
measurement.}
\label{fig:3}
\end{figure}

We focus here on the decays of $K_L$ which do not suffer from the uncertainties related to the isospin corrections that are present in the decays of charged kaons. These being 3-body decays, 
they are not helicity suppressed and, therefore, we do not expect a strong deviation from the SM prediction after experimental constraints are taken into account. While the non-unitarity effect and 
the modification of the phase-space due to the presence of a heavy sterile neutrino if it is kinematically accessible are always present, the stringent limits on leptonic mixing strongly limit the size of the 
allowed deviations. This is evident from Fig.~\ref{fig:3} (left)
where we present our predictions for the decays $K_L \rightarrow \pi^- e^+ \nu$. We checked as well that no sizeable deviations are present in the lepton polarization asymmetry and in the 
forward-backward asymmetry.

\subsection{Loop-induced weak decay $K\to \pi \nu \nu$}

The semileptonic decays $K\to \pi \nu \nu$ is of particular interest since it is dominated by short-distance contributions which arise at the one-loop level. As a consequence, this process is
especially sensitive to the presence of new heavy particles in the loop. Thus a precise measurement of these decays would provide new constraints or could point towards the existence of new physics.
It is worth noting that a control over the remaining long-distance hadronic contribution to the charged mode has recently been achieved~\cite{Christ:2016eae}, allowing for a reduced theoretical
uncertainty. Both neutral and charged decays are also subjects of intense experimental searches at NA62~\cite{Hahn:1404985,Volpe:2017lyp} and KOTO~\cite{Ahn:2016kja}. Our predictions for 
the branching ratio of $K^+\to \pi^+ \nu \nu$ can be found in Fig.~\ref{fig:3} (right). While a heavy neutrino can induce percent-level deviations, the theoretical uncertainties on the SM 
predictions are at the $10\%$ level. We obtained similar results for $K_L\to \pi^0 \nu \nu$ , which effectively makes these decays blind to the presence of an extra sterile neutrino.

\subsection{``Invisible decay" $K_L\to \nu\nu$}

\begin{figure}[t!]
\centering
\includegraphics[width=0.65\linewidth]{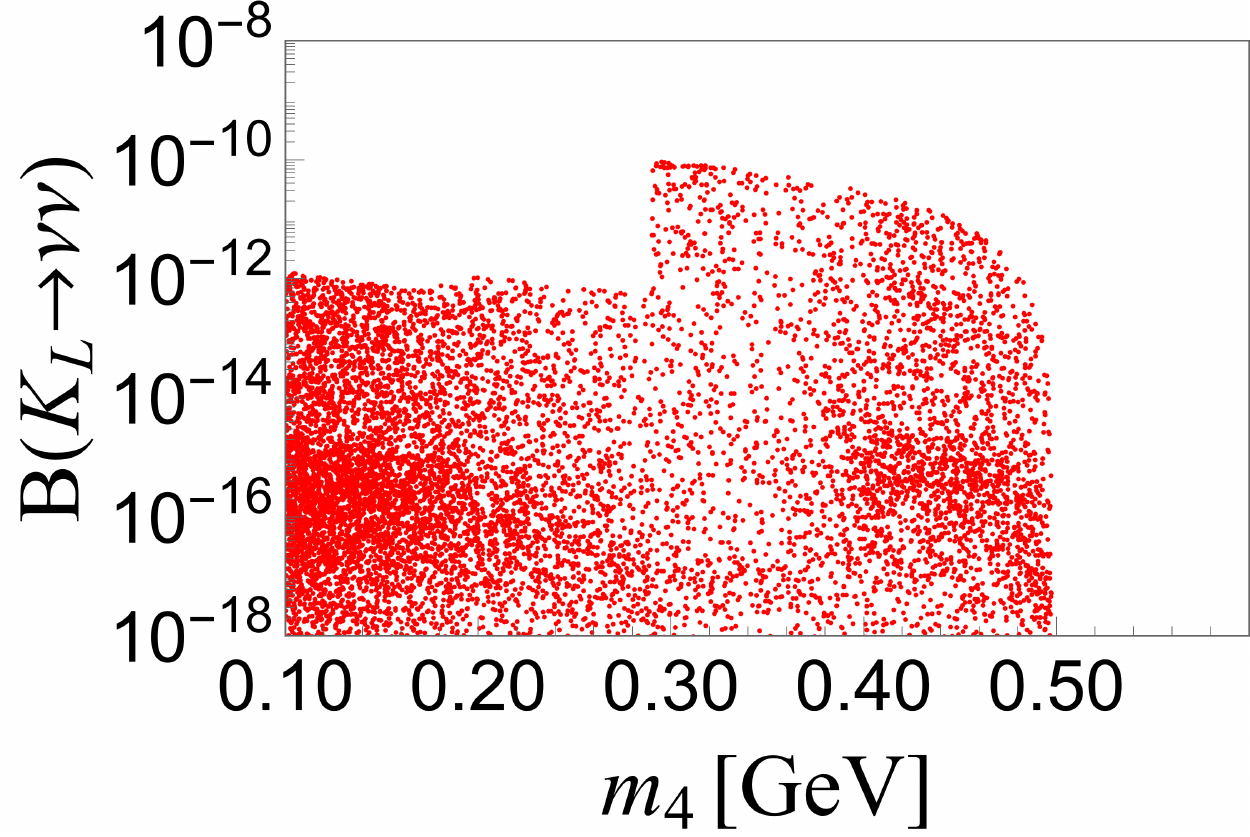}
\caption{\small \sl $ \mathrm{Br}(K_L\to \nu \nu) $ as a function of the mass of the sterile neutrino $m_4$. The points presented pass all the constraints considered in this study.}
\label{fig:4}
\end{figure}

The last process considered in this work is the decay $K_L\to \nu\nu$. In the SM, the branching ratio of this helicity-suppressed decay is exactly zero with massless neutrinos. When massive 
neutrinos are considered, we get the following result
\begin{align}\label{Knunu}
\mathrm{Br} (K_L\rightarrow \nu\nu)& = \mathop{ \sum_{i,j=1}^{4}}_{i\leq j}  \left( 1 - \frac{1}{2 }\delta_{ij} \right)  \frac{\alpha_{\rm em}^2 G_F^2 \tau_{K_L}}{8 \pi^3 m_{K}^3 \sin^4\theta_W} f_K^2 \lambda^{1/2}(m_{K}^2,m_{\nu_i}^2,m_{\nu_j}^2)\\
& \times \left[  \biggl| \sum_{\ell\in \{e,\mu,\tau\}} \!\! \mathrm{Re}\left( \lambda_c X_c^\ell + \lambda_t X_t \right) U_{\ell i}^\ast U_{\ell j}\biggr|^2 \left( m_K^2 (m_{\nu_i}^2 + m_{\nu_j}^2 ) - (m_{\nu_i}^2 - m_{\nu_j}^2)^2 \right) \right. \nonumber\\
& \left.+\ 2\!\! \sum_{\ell,\ell'\in \{e,\mu,\tau\}} \!\! \mathrm{Re}( \lambda_c X_c^\ell + \lambda_t X_t ) \mathrm{Re}( \lambda_c X_c^{\ell'} + \lambda_t X_t  ) \mathrm{Re}( U_{\ell i}^\ast U_{\ell j} U_{\ell' i}^\ast U_{\ell' j} ) \; m_{\nu_i}  m_{\nu_j} m_K^2\right]\,,\nonumber
\end{align}
where, in addition to the quantities previously defined, we have $\lambda_c = V_{cs}^\ast V_{cd}$, $\lambda_t = V_{ts}^\ast V_{td}$ and the quark loop functions giving $X_t=1.47(2)$ for the top
contribution~\cite{Brod:2010hi} and $X_c^e = X_c^\mu = 10.0(7)\times 10^{-4}$, $ X_c^\tau = 6.5(6)\times 10^{-4}$ for the charm contributions~\cite{Buras:1998raa}. In the presence 
of only 3 light massive neutrinos with masses and mixing in agreement with oscillation data, we get an extremely suppressed prediction $\mathrm{Br} (K_L\rightarrow \nu\nu)<10^{-20}$.
As such an observation
of this decay with a branching ratio larger by a few orders of magnitude or more would be a clear signal of new physics. Our predictions are presented in Fig.~\ref{fig:4}
where we see that that a heavy neutrino could increase $\mathrm{Br} (K_L\rightarrow \nu\nu)$ up to $1.2\times10^{-10}$. Unfortunately this falls short of the of NA64 expected sensitivity to this
decay~\cite{Gninenko:2015mea} where this decay could be searched for by using $K_L$ produced from a $K^+$ beam hitting an active target. This motivates the study of other set-ups which
could be sensitive to this decay or similar invisible decays. In any case, an experimental bound on this decay mode would be of great importance for studying physics beyond the SM.

\section{Conclusion}

In this work, we used a $3+1$ model to describe the effects of a heavy sterile neutrino on leptonic and semileptonic kaon decays, providing a simplified framework that should
reproduce the effects of a more realistic model where the SM is extended to include one or more sterile neutrinos and reproduce the low-energy neutrino oscillation data.
We have focused on a heavy neutrino with a mass close to the kaon mass and studied in particular the effect of a kinematically accessible sterile neutrino. We found that sizeable deviations
from the SM prediction and experimental measurement are expected in $K_{e2}$ and lepton universality tests, allowing to derive new constraints on the mass and mixing of the sterile neutrino.
We also derived the expressions for the semileptonic kaon decays $\mathrm{Br}(K \to \pi \nu\nu)$ and
$\mathrm{Br}(K_L \to \nu\nu)$ in the case of massive neutrinos and provide generic results that can be used
when studying new physics scenarios in which sterile heavy neutrinos present. Here, the deviations due to a heavy neutrino are much smaller than the experimental or theoretical
uncertainties. Finally, we considered the decay $K_L \to \nu\nu$ which would be a smoking gun of new physics if observed due to the extremely suppressed SM prediction, smaller than $10^{-20}$. The presence of a heavy neutrino
with a mass below $m_K$ could increase $\mathrm{Br}(K_L \to \nu\nu)$ up to $\mathcal{O}( 10^{-10})$, a value much above the SM prediction and just a couple orders of magnitude below the expected
sensitivity of NA64. This motivates the study of different experimental set-ups that could make use of the large number of kaons produced in fixed target experiments or in low-energy colliders 
like DAFNE.

\section*{Acknowledgements}

This project has received funding from the European Union's Horizon 2020 research and innovation program under the Marie Sklodowska-Curie Grants No. 690575 and No. 674896.
C.W. receives financial support from the European Research Council under the European Union's Seventh Framework Programme
(Grants No. FP/2007-2013)/ERC Grant NuMass Grant No. 617143. He also wishes to acknowledge the Moriond organizing committee for the financial support that allowed him to attend the conference.


\end{document}